\documentclass[aps,reprint,showpacs,showkeys,preprintnumbers,groupedaddress,amsmath,amssymb,]{revtex4-1}

\usepackage{graphicx}% Include figure files
\usepackage{dcolumn}% Align table columns on decimal point
\usepackage{bm}% bold math
\usepackage{hyperref}% add hypertext capabilities
\usepackage[dvips]{color}

\begin{document}


%\preprint{}


\title{Electron effective mass and electronic structure in nonstoichiometric a-IGZO films}


\author{Xin Zhu}
\affiliation{Tianjin Key Laboratory of Low Dimensional Materials Physics and
Preparing Technology, Department of Physics, Tianjin University, Tianjin 300354,
China}
\author{Yang Yang}
\affiliation{Tianjin Key Laboratory of Low Dimensional Materials Physics and
Preparing Technology, Department of Physics, Tianjin University, Tianjin 300354,
China}
\author{Xin-Dian Liu}
\affiliation{Tianjin Key Laboratory of Low Dimensional Materials Physics and
Preparing Technology, Department of Physics, Tianjin University, Tianjin 300354,
China}
\author{Zhi-Qing Li}
\email[Author to whom correspondence should be addressed. Electronic address: ]{zhiqingli@tju.edu.cn}
\affiliation{Tianjin Key Laboratory of Low Dimensional Materials Physics and
Preparing Technology, Department of Physics, Tianjin University, Tianjin 300354,
China}

%\thanks{}

\date{\today}

\begin{abstract}
The transport properties and optical transmittance and absorption spectra for the  nostoichiometric amorphous Indium Gallium Zinc Oxide (a-IGZO) films with Gallium and Zinc deficiencies are investigated.
The resistivity and carrier concentration variation with temperature both reveal that the films possess degenerate semiconductor (or metal) characteristics.
The thermopower is negative and decreases linearly with decreasing temperature, indicating the electron diffusion thermopower governs the thermal transport process in each film. Using free-electron-like model, we extracted the electron effective mass, which is about three times as large as that of the stoichiometric one and increases with increasing carrier (electron) concentration. Neglecting the variation in the energy with the wavevector near the valence band maximum and using the free-electron-like model, we also obtained the electron effective mass via the optical absorption spectra measurement.
The magnitude of the effective mass obtained via optical spectra measurement is  comparable to that obtained via thermopower measurement for each film. Our results strongly suggest that the nostoichiometric a-IGZO films possess free-electron-like pseudo-energy-bandstructure.
\end{abstract}

%\keywords{}

\maketitle

%\section{}
%\section{Introduction\label{sec1}}
%\subsection{}
%\subsubsection{}
\section{Introduction}
Since the the report on transparent and flexible thin-film transistors (TFTs) using amorphous In-Ga-Zn oxide (a-IGZO) as active channel material~\cite{Nature-2004-Honoso}, extensive investigations have been carried out on this material both in device fabrications~\cite{APL-2009-Nomura, APL-2018-Jallorina, Thin solid films-2017-Yang, JAC-2017-Yang, JPD;Ap-2017-Na, STAM-2010-Kamiya} and fundamental research~\cite{STAM-2010-Kamiya, PRB-2007-Nomura, PSSA-2009-Kamiya, PRB-2010-Medvedeva, JP:CM-2017-Zhou, JCE-2017-Meux, PRB-2011-Noh, PRA-2018-Meux}.
The electronic mobility in a-IGZO based TFT is far larger than that in amorphous Si:H based device~\cite{Nature-2004-Honoso, STAM-2010-Kamiya}. While compared to the polycrystalline transparent oxide, such as ZnO, and SnO$_2$, a-IGZO has excellent uniformity (no grain boundaries) and chemical stability. In addition, it is found that the mobility of a-IGZO can be further increased by annealing in wet O$_2$ atmosphere~\cite{APL-2009-Nomura, APL-2018-Jallorina}, or increasing  O$_2$ ratio (the ratio of O$_2$ to Ar)~\cite{Thin solid films-2017-Yang} in sputtering process. The stability of a-IGZO TFTs could be further enhanced by introducing gate insulator layer, such as Al$_2$O$_3$~\cite{JAC-2017-Yang} and HfO$_2$~\cite{JPD;Ap-2017-Na}.  The relative high mobility of a-IGZO closely relates to its electronic structure. The pseudo-bandstructures of a-IGZO are similar to the energy-bandstructures of the crystalline IGZO: both materials are direct-gap semiconductor, the conduction band bottoms of both materials are mainly formed by the 5\emph{s} orbital of In$^{3+}$, and the overlaps between the 5\emph{s} wave function of In$^{3+}$ ions are not altered significantly by the disordered local structures in a-IGZO~\cite{PRB-2007-Nomura, PSSA-2009-Kamiya, PRB-2010-Medvedeva, JP:CM-2017-Zhou, JCE-2017-Meux}.

The electron effective mass is an important quantity to affect the electron mobility, which is crucial to the performance of the TFTs. According to previous reports, the electron effective mass $m^{*}$ of the a-InGaZnO$_4$ is $m^{*}\simeq0.34m_e$ ($m_e$ is the free electron mass)~\cite{Thin solid films-2005-Takagi, Thin solid films-2010-Hosono-0.34}, which is similar to  that of single crystal InGaZnO$_4$~\cite{APL-2004-Nomura, Thin solid films-2005-Takagi}. In practice, the a-IGZO films are often fabricated by rf magnetron sputtering method. For this method, the atomic fraction of elements is sensitive to the depositing condition and the deviation to the stoichiometric ratio for the metal ions is inevitable, which will alter the electronic structure and electron effective mass of a-IGZO films. Thus it is necessary to investigate the electronic structure and electron effective mass of nonstoichiometric a-IGZO films prepared using a stoichiometric IGZO target.

In the present paper, we measured the temperature dependence of resistivities and thermoelectric powers (thermopowers) of several nostoichiometric a-IGZO films. The results can be well explained by the free-electron-like model. Thus we obtain the electron effective mass of each film via both the temperature dependence of thermopower data and optical band gap measurement. It is found that the electron effective mass of the film with Ga and Zn deficiencies is much larger than that of the stoichiometric one, and the electron effective mass increases with increasing the carrier concentration.

\section{Experimental Method}
The a-IGZO films were deposited on glass substrates (Fisherfinest premium microscope slide) by standard rf sputtering method. The sputtering source was a commercial InGaZnO$_4$ ceramic target with purity of $99.99$\% and atomic ratio of In, Ga, and Zn being $1:1:1$.  Base
pressure below $1 \times10^{-4}$\ Pa was established prior to each deposition run. The sputtering was carried out in an argon ($99.999$\% in purity) atmosphere with the pressure of 0.6\,Pa, and the sputtering power was maintained at 70\,W during the deposition progress. To obtain films with different carrier concentration, the substrate temperatures $T_s$ were set as 295, 405, 535 and 685\,K, respectively, and the corresponding films are designated as No.1, No.2, No.3, and No.4, respectively.

The thicknesses of the films ($\sim$1\,$\mu$m) were measured using a surface profiler (Dektak, 6M). The crystal structures of the films were  measured
in a powder X-ray diffractometer (D/MAX-2500, Rigaku) with Cu K$_\alpha$ radiation. The atomic fraction of metallic elements of each film was obtained by energy dispersive X-ray spectroscopy (EDX) and listed in Table~\ref{Table1}. Optical transmittance and absorption spectra were measured in the wavelength range of 300-800 nm by
using an UV-vis spectrometer (Hitachi V4100).  The resistivities and Hall effect measurements were carried out in a physical property measurement system (PPMS-6000, Quantum Design) by employing the standard four-probe method. The Al electrodes were deposited on the film to obtain good contacts.

\begin{figure}
\begin{center}
\includegraphics[scale=0.9]{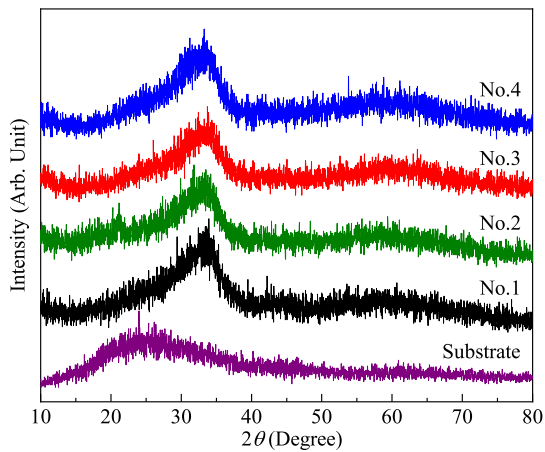}
\caption{(Color online) XRD patterns of the a-IGZO films deposited at different substrate temperatures. The pattern in the bottom is the diffraction pattern of  the glass substrate.}\label{FigXRD}
\end{center}
\end{figure}

\section{Results and Discussions}
Figure~\ref{FigXRD} shows X-ray diffraction patterns of the films. The diffraction pattern of the glass substrate is also presented for comparison. For each film, there are two halo peaks centered around 34$^{\circ}$ and 58$^{\circ}$, which are typical for the a-IGZO films~\cite{Nature-2004-Honoso, Thin solid films-2005-Takagi}. Thus our films are amorphous even if the substrate temperature is as high as 685\,K. The EDS results are summarized in Table~\ref{Table1}, from which one can see that both Ga and Zn atoms are less than the stoichiometric ratios and the atomic fraction ratios of Ga and Zn slightly decreases with increasing substrate temperature.

\begin{figure}
\begin{center}
\includegraphics[scale=0.9]{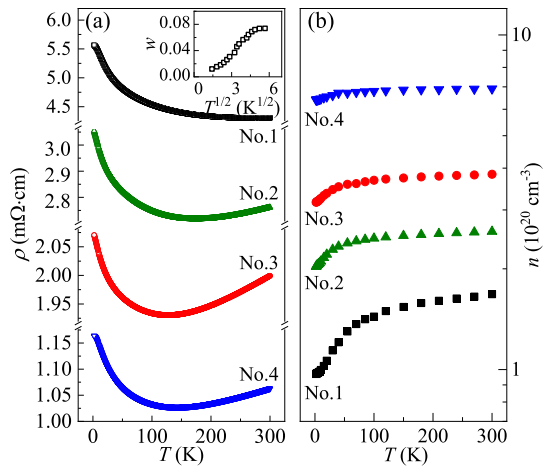}
\caption{(Color online) (a) Resistivity versus temperature for the four a-IGZO films. (b) The carrier concentration (electron density) obtained via Hall effect measurements versus temperature for the a-IGZO films. The inset of (a) shows $w=\mathrm{d}\ln{\sigma}/\mathrm{d}\ln{T}$ versus $T^{1/2}$ for film No.1.}\label{FIGRTNT}
\end{center}
\end{figure}

Figure~\ref{FIGRTNT}(a) shows the resistivity as a function of temperature from 300 down to 2\,K. The temperature coefficients of resistivity TCR ($\text{TCR}=\rm{d}\rho/\rho \rm{d}T$) for films No.2, No.3, and No.4 are positive above $\sim$150\,K. In addition, the carrier concentrations (electron density) of these films are insensitive to the temperature above this temperature [Fig.~\ref{FIGRTNT}(b)]. These features mean films No.2 to No.4 all possess degenerate semiconductor (or metal) characteristics in electrical transport properties. The enhancement of resistivity with decreasing temperature below $\sim$150\,K originates from weak localization effect\cite{RMP-1985-Lee, JAP-2013-Shinozaki, Thin solid films-2014-Shinozaki}  and electron-electron interaction effect~\cite{Thin solid films-2014-Shinozaki, PRB-1980-Altshler, PRB-2003-Efetov, PRL-2003-Belo, PRB-2015-Wu, PSSB-2017-Zhang}, while the reduction of carrier concentration is also caused by the electron-electron interaction effect~\cite{PRB-2015-Wu, PSSB-2017-Zhang, PRL-2007-Kharitonov, PRB-2008-Kharitonov}.
According to  M\"{o}bius~\cite{PRB-1989-Mobius-w, CRSS-2017-Mobius}, the logarithmic derivative of
the conductivity $w=\mathrm{d} \ln\sigma/ \mathrm{d} \ln T$ is more sensitive than the TCR and defines a more accurate and reliable criterion to distinguish between metallic
and insulating behaviors. For film No.1 the value of $w$ tends to be zero as $T$  approaching to zero [see inset of Fig. ~\ref{FIGRTNT}(a)] and the carrier concentration also keeps as a constant from $\sim$100\, to 300\,K, which means this film also possesses degenerate semiconductor characteristics in transport properties.

\begin{table*}
\caption{\label{Table1} Relevant parameters for the four a-IGZO films. Here $T_s$ is the substrate temperature during deposition, $n(300\,\mathrm{K})$ is carrier concentration (electron density) at 300\,K obtained via Hall effect measurement.  $E_F$ and $m_s^*$ are Fermi energy and electron effective mass obtained via thermopower measurements, respectively. $E_g$ is the optical band gap and $m_o^*$ is the electron effective mass obtain via Eq.~(\ref{Eq-Eg2}).}
\begin{ruledtabular}
\begin{center}
\begin{tabular}{ccccccccc}%\hline \hline
        &$T_s$  &   In$:$Ga$:$Zn          & $\rho(300\,\mathrm{K})$ & $n(300\,\mathrm{K})$   &  $E_F$     &  $m_s^*$ &  $E_g$ & $m_o^*$   \\
Film    & (K)   &   (Atomic ratio)        & (m$\Omega$\,cm)          & ($10^{20}$cm$^{-3}$)   &   (meV)    &  ($m_e$) &  (eV)  & ($m_e$)    \\ \hline

No.1       &295    &  $1:0.87:0.57$          &4.31                         &1.68                     &119.80      &0.93     &3.57      &0.74     \\
No.2       &405    &  $1:0.85:0.57$          &2.76                         &2.59                     &149.06      &0.99     &3.62      &0.74     \\
No.3       &535    &  $1:0.85:0.54$          &2.00                         &3.84                     &189.71      &1.02     &3.66      &0.80     \\
No.4       &685    &  $1:0.84:0.43$          &1.06                         &6.91                     &255.34      &1.12     &3.71      &0.86     \\%\hline \hline
\end{tabular}
\end{center}
\end{ruledtabular}
\end{table*}

\begin{figure}
\begin{center}
\includegraphics[scale=0.8]{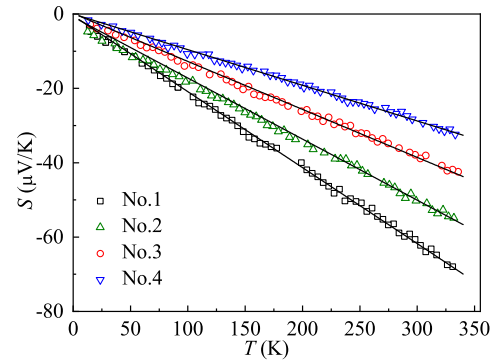}
\caption{(Color online) The thermopowers $S$ versus temperature for the four a-IGZO films deposited at different substrate temperature. The straight solid lines are the least-squares fits to Eq.~(\ref{EqST}).} \label{FIGST}
\end{center}
\end{figure}

Figure~\ref{FIGST} displays the thermopowers variation with temperature from 330 down to 15\,K for the four films, as indicated. One can clearly see that the thermopower of each film is negative and varies linearly with temperature in the whole measured temperature range. The negative thermpower means the main charge carrier is electron instead of hole, which is agreed with the result obtained from Hall effect measurement.
 In a typical metal, the  thermopower $S$ is generally composed of two parts: the electron diffusion thermopower $S_d$ and the phonon-drag thermopower $S_g$. The former originates from the diffusion of electrons from the hotter end to the colder side of the sample, while the latter arises from the electron-phonon interaction.  According to free-electron model, the diffusion thermopower is proportional to the temperature $T$ at $T\ll\theta_D$ ($\theta_D$ being the Debye temperature) and can be written as~\cite{Wiley-1962-MacDonald}
\begin{equation}\label{EqST}
S_d=-\frac{\pi^2 k_B T}{3 |e| E_F},
\end{equation}
where $k_B$ is Boltzmann constant, $e$ is the electron charge, and $E_F$ is the Fermi energy. The
phonon-drag thermopower is nonlinearly dependent with the temperature and dominates in the clean samples with long phonon relaxation
time. In the presence of high level of disorder, the phonon-drag contribution would be suppressed and negligible. Since long-range order is absent in amorphous material, it is expected that the magnitudes of the phonon-drag thermopowers of the a-IGZO films are too small to be negligible, which is just the case shown in Fig.~\ref{FIGST}. In fact, the phono-drag thermopowers in Sn doped In$_2$O$_3$ (ITO) films~\cite{JAP-2004-Li-ITO, JAP-2010-Wu-ITO} and F doped SnO$_2$ (FTO) films~\cite{APL-2014-Lang-FTO} are also found too weak to be ignored. It has been found that the thermopowers of both the ITO films and FTO films vary linearly with temperature from $\sim$300\,K down to the liquid helium temperatures, and the carrier concentrations of the films obtained from the temperature dependence of thermopowers using free-electron-like model are consistent with that obtained from Hall effect measurements within the experimental uncertainties~\cite{JAP-2004-Li-ITO, JAP-2010-Wu-ITO, APL-2014-Lang-FTO}.

The linear temperature dependent behavior of $S(T)$ for a-IGZO films has already indicated that the conduction electrons in the films possess free-electron-like characteristics. Considering the Debye temperature of the transparent conducting oxide material is far larger than the room temperature~\cite{JAP-2004-Li-ITO, JAP-2010-Wu-ITO, APL-2014-Lang-FTO, Nanotec-2009-Chiu, ASS-2012-Liu, PRB-2013-Preissler}, we compare the experimental $S(T)$ data with Eq.~(\ref{EqST}). The slope of the temperature dependence $S(T)$ data is obtained by least-square method, then the value of the Fermi energy $E_F$ can be obtained. Using the free-electron-like model, i.e., $E_F=\hbar^2k_F^2/2m^{*}$ and $k_F=(3\pi^2n)^{1/3}$ with $\hbar$ being the Planck's constant divided by $2\pi$, $k_F$ the magnitude of the Fermi wavevector, and $n$ the carrier concentration, one can readily obtain the relation $m^{*}=\hbar^2(3\pi^2 n)^{2/3}/2E_F$. The value of $n$ can be obtained via Hall effect measurement, thus the electron effective mass of each film is obtained and listed in Table~\ref{Table1} (denoted by $m_s^*$). Inspection of Table~\ref{Table1} indicates that the electron effective mass in the nonstoichiometric a-IGZO films varies from $\sim$$0.93m_e$ to $\sim$$1.12m_e$ and increases with the enhancement of carrier concentration. The magnitude of the electron effective mass of nonstoichiometric a-IGZO films is about three times as large as that of the stoichiometric ones  (The effective mass of the stoichiometric a-IGZO$_4$ is $\sim$0.34\,$m_e$~\cite{Thin solid films-2005-Takagi, Thin solid films-2010-Hosono-0.34}), which indicates that the curvature of the energy-wavevector dispersion curve near the bottom of the conduction band of the former is less than that of the latter. In degenerated In$_2$O$_3$ materials, it is also found that the electron effective mass increases rapidly with increasing electron density~\cite{PRB-2008-Fuchs}. Thus this phenomena may be universal in the transparent conducting oxide materials.

\begin{figure}
\begin{center}
\includegraphics[scale=0.75]{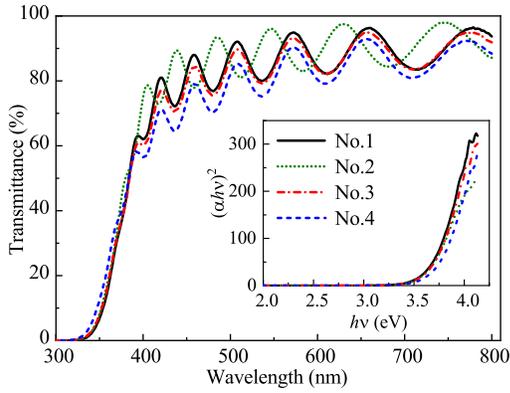}
\caption{(Color online) Optical transmission spectra of a-IGZO films deposited at different substrate temperatures. Inset: the relationship between $(\alpha h\nu)^2$ and $h\nu$ for these films.} \label{FIGOPT}
\end{center}
\end{figure}

To test the validity of the results obtained from transport measurements, we also measured the optical transmittance spectra of the a-IGZO films, and the results are presented in Fig.~\ref{FIGOPT}, in which the transmittance $T_R$ is the intensity ratio of transmitted light ($I$) to incident light ($I_0$). Clearly, all films exhibit relatively high transparency in the visible range (380-780\,nm), and the transmittance tend to decrease with increasing electron density. As mentioned above, the density functional theory (DFT) calculation results indicate the pseudo-bandstructure~\cite{PRB-2007-Nomura, PSSA-2009-Kamiya, PRB-2010-Medvedeva, JP:CM-2017-Zhou, JCE-2017-Meux} of the stoichiometric amorphous InGaZnO$_4$ is similar to that of the crystalline one~\cite{Thin solid films-2010-Hosono-0.34}, and both of the materials are direct gap semiconductor. For the direct gap transition, the relationship between absorption coefficient $\alpha$ and the incident photon energy can be written as~\cite{PSSB-1966-Tauc, Prentice-1971-Pankove}
\begin{equation}\label{Eq-Absorb}
(\alpha h\nu)^2=A(h\nu-E_g),
\end{equation}
where $h$ is Planck's constant, $A$ is the energy independent constant, $\nu$ is the frequency of the incident photon, and $E_g$ is the optical band gap. Using $I=I_0\exp(-\alpha d)$, with $d$ being the thickness of the film one can readily obtain the relation between $(\alpha h\nu)^2$ and $h\nu$, which was presented in the inset of Fig.~\ref{FIGOPT}. From the inset of Fig.~\ref{FIGOPT}, we obtain the value of $E_g$ by extrapolating the linear part of the plot to $(\alpha h\nu)^2=0$. The obtained values of $E_g$ are summarized in Table~\ref{Table1}.

\begin{figure}
\begin{center}
\includegraphics[scale=1.1]{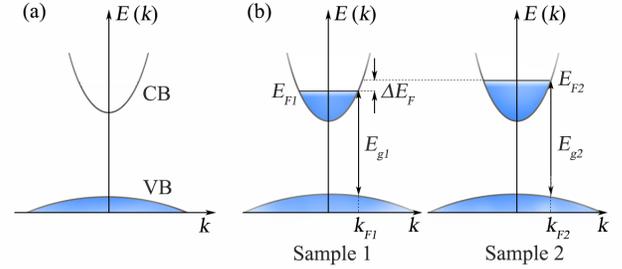}
\caption{Heuristic energy bandstructure near the top of the valence band and the bottom of conduction, (a) for the stoichiometric  amorphous InGaZnO$_4$,  and (b) for two nonstoichiometric a-IGZO samples with different carrier concentrations. Here CB and VB  are the abbreviations of conduction band and valence band, respectively.}\label{Fig-Bandstructure}
\end{center}
\end{figure}

Combing the DFT calculation results~\cite{PRB-2007-Nomura, JP:CM-2017-Zhou, PRB-2007-Nomura, PSSA-2009-Kamiya, JCE-2017-Meux, PRB-2010-Medvedeva} and the transport properties discussed above, we give a schematic pseudo-bandstructure of the nonstoichiometric a-IGZO film (Fig.~\ref{Fig-Bandstructure}). The variation of energy $E$ with wavevector $k$ is slow near the maximum of the conduction band, i.e., the top of the valence band is nearly flat, and the energy-wave vector dispersion relation in the vicinity of the conduction band minimum is parabolic.  The Fermi level lies in the gap for the stoichiometric amorphous InGaZnO$_4$, while it move up to the conduction band due to the introduction of defects for the nonstoichiometric a-IGZO films~\cite{STAM-2010-Kamiya, PSSA-2009-Kamiya}. Neglecting the energy change with wavevector near the top of the valence band, one can obtain the change of the band gap
\begin{equation}\label{Eq-Eg1}
\Delta{E_g}\approx \Delta{E_F}=E_{F2}-E_{F1},
\end{equation}
for two samples with Fermi energy $E_{F1}$ and $E_{F2}$. Here we also use the free-electron-like model, and Eq.~(\ref{Eq-Eg1}) can be written as
\begin{equation}\label{Eq-Eg2}
\Delta{E_g}\approx \frac{(3{\pi^2}{\hbar^3}{n_2})^{\frac{2}{3}}}{2{m_2}^*}-\frac{(3{\pi^2}{\hbar^3}{n_1})^{\frac{2}{3}}}{2{m_1}^*},
\end{equation}
where $n_1$ ($n_2$) and $m_1^*$ ($m_2^*$) are the electron density and electron effective mass for sample 1 (sample 2), respectively.

For the nonstoichiometric a-IGZO  films, we firstly assume the electron effect mass of film No.1 is identical to that of film No.2, then the electron effective masses in films No.1 and No.2 can be obtained, $m_1^*=m_2^*=0.74m_e$.  Substituting this value into Eq.~(\ref{Eq-Eg2}), one can readily obtain the electron effective mass in film No.3. The electron effective mass obtained via Eq.~(\ref{Eq-Eg2}) for each film is also listed in Table~\ref{Table1} and designated as $m_o^*$. Again, the magnitude of the effective mass increases with increasing electron density, and is much larger than that for the stoichiometric amorphous InGaZnO$_4$ films.
In addition, the $m_o^*$ value is about $80\%$ of $m_s^*$ for each film. We notice that the slight difference between
$m_o^*$ and $m_s^*$  is not unexpected. Such discrepancy may be caused by two reasons: (i) for the a-IGZO  films, besides the cations \emph{s} states, the the oxygen \emph{p} states also have a little contribution to the conduction band~\cite{PRB-2010-Medvedeva}, which means that the dispersion relation between energy and wavevector near the conduction band minimum would not be strictly parabolic and deviation could be generate in using the free-electron-like model; (ii) the diffusion thermopower in Eq.~(\ref{EqST}) can be more accurately expressed as $S_d=[-\pi^2k_BT/(3|e|E_F)]\xi$, in which the correction parameter $\xi$ may somewhat deviate from unity in real metals or degenerate semiconductors~\cite{Wiley-1972-Barnad}. In the end, we should emphasize that the magnitudes of the electron effective mass obtained via thermopower and optical absorbtion coefficient measurements are comparable for each nonstoichiometric a-IGZO film, which in turn indicates that the pseudo-bandstructure of the nonstoichiometric a-IGZO films can be well described by the heuristic model in Fig.~\ref{Fig-Bandstructure}.

\section{Conclusion}
In summary, we have fabricated a series of a-IGZO films by rf sputtering method using a stoichiometric InGaZnO$_4$ target. It was found both Ga and Zn cations are less than the  stoichiometric ratio.
The temperature dependence of resistivity indicates the films are degenerate semiconductors, which is confirmed by the temperature dependence of Hall effect measurements. The thermopower is negative and its absolute value decreases linearly with decreasing temperature, which means that electron diffusion mechanism dominant the thermal transport process and the contribution of phonon-drag effect is negligible. Using free-electron-like model, we extracted the electron effective masses of the nonstoichiometric films and found the values are about three times as large as that of the stoichiometric ones and increase with increasing electron density.
Basing the free-electron-like model, we also obtained the effective mass through optical absorption spectra measurements. It is found that the magnitudes of the effective mass obtained independently by the two different methods are comparable for each film. Our results strongly suggest that the nonstoichiometric a-IGZO films possess free-electron-like pseudo-energy-bandstructure.

\begin{acknowledgments}
This work is supported by the National Natural Science Foundation of China through Grant No. 11774253.
\end{acknowledgments}


\begin{thebibliography}{00}\label{sec:TeXbooks}
\bibitem{Nature-2004-Honoso}K. Nomura, H. Ohta, A. Takagi, T. Kamiya, M. Hirano, and H. Hosono, Nature \textbf{432}, 488 (2004).
\bibitem{APL-2009-Nomura}K. Nomura, T. Kamiya, M. Hirano, and H. Hosono, Appl. Phys. Lett. \textbf{95}, 013502 (2009).
\bibitem{APL-2018-Jallorina}M. P. A. Jallorina, J. P. S. Bermundo, M. N. Fujii, Y. Ishikawa, and Y. Uraoka, Appl. Phys. Lett. \textbf{112}, 193501 (2018).
\bibitem{Thin solid films-2017-Yang}D. G. Yang, H. D. Kim, J. H. Kim, S. W. Lee, J. Park, Y. J. Kim, and H. S. Kim, Thin Solid Films \textbf{638}, 361 (2017).
\bibitem{JAC-2017-Yang}D. G. Yang, H. D. Kim, J. H. Kim, K. Park, J. H. Kim, Y. J. Kim, J. Park, and H. S. Kim, J. Alloy. Compd. \textbf{729}, 1195 (2017).

\bibitem{JPD;Ap-2017-Na}S. Y. Na, Y. M. Kim, D. J. Yoon and S. M. Yoon, J. Phys. D: Appl. Phys. \textbf{50}, 495109 (2017).
\bibitem{STAM-2010-Kamiya}T. Kamiya, K. Nomura and H. Hosono, Sci. Technol. Adv. Mater. \textbf{11}, 044305 (2010).
\bibitem{PRB-2007-Nomura}K. Nomura, T. Kamiya, H. Ohta, T. Uruga, M. Hirano, and H. Hosono, Phys. Rev. B \textbf{75}, 035212 (2007).
\bibitem{PSSA-2009-Kamiya}T. Kamiya, K. Nomura, and H. Hosono, Phys. Status Solidi A \textbf{206}, 860 (2009).
\bibitem{PRB-2010-Medvedeva}J. E. Medvedeva, and C. L. Hettiarachchi, Phys. Rev. B \textbf{81}, 125116 (2010).

\bibitem{JCE-2017-Meux}A de Jamblinne de Meux, G. Pourtois, J. Genoe, and P. Heremans, J. Phys.: Condens. Matter \textbf{29}, 255702 (2017).
\bibitem{JP:CM-2017-Zhou}X. L. Zhou, H. X. Cao, Z.B. Zhou, J. C. Cao, and J. Yu, J. Comput. Electron. \textbf{16}, 280 (2017).
\bibitem{PRB-2011-Noh}H. K. Noh and K. J. Chang, Phys. Rev. B \textbf{84}, 115205 (2011).
\bibitem{PRA-2018-Meux}A de Jamblinne de Meux, G. Pourtois, J. Genoe, and P. Heremans, Phys. Rev. Appl. \textbf{9}, 054039 (2018).
\bibitem{Thin solid films-2010-Hosono-0.34}Y. Kikuchi, K. Nomura, H. Yanagi, T. Kamiya, M. Hirano, and H. Hosono, Thin Solid Films \textbf{518}, 3017 (2010).

\bibitem{Thin solid films-2005-Takagi}A. Takagi, K. Nomura, H. Ohta, H. Yanagi, T. Kamiya, M. Hirano, and H. Hosono, Thin Solid Films \textbf{486}, 38 (2005).
\bibitem{APL-2004-Nomura}K. Nomura, T. Kamiya, H. Ohta, K. Ueda, M. Hirano, and H. Honoso, Appl. Phys. Lett. \textbf{85}, 1993 (2004).
\bibitem{RMP-1985-Lee}P. A. Lee and T. V. Ramakrishnan, Rev. Mod. Phys. \textbf{57}, 287 (1985).
\bibitem{JAP-2013-Shinozaki}B. Shinozaki, K. Hidaka, S. Ezaki, K. Makise, T. Asano, S. Tomai, K. Yano, and H. Nakamura, J. Appl. Phys. \textbf{113}, 153707 (2013).
\bibitem{Thin solid films-2014-Shinozaki}B. Shinozaki, K. Hidaka, S. Ezaki, K. Makise, T. Asano, S. Tomai, K. Yano, and H. Nakamura, Thin Solid Films \textbf{551}, 195 (2014).

\bibitem{PRB-1980-Altshler}B. L. Altshler, D. Khmel'nitzkii, A. I. Larkin, and P. A. Lee,  Phys. Rev. B \textbf{22}, 5142 (1980).
\bibitem{PRB-2003-Efetov}K. B. Efetov and A. Tschersich, Phys. Rev. B \textbf{67}, 174205 (2003).
\bibitem{PRL-2003-Belo}I. S. Beloborodov, K. B. Efetov, A.V. Lopatin, and V. M. Vinokur, Phys. Rev. Lett. \textbf{91}, 246801 (2003).
\bibitem{PRB-2015-Wu}Y. N. Wu, Y. F. Wei, and Z. Q. Li, Phys. Rev. B. \textbf{91}, 104201 (2015).
\bibitem{PSSB-2017-Zhang}H. Zhang, X. J. Xie, X. H. Zhang, X. D. Liu, and Z. Q. Li, Phys. Status Solidi B \textbf{254}, 1700133 (2017).

\bibitem{PRL-2007-Kharitonov}M. Y. Kharitonov and K. B. Efetov, Phys. Rev. Lett. \textbf{99}, 056803 (2007).
\bibitem{PRB-2008-Kharitonov}M. Y. Kharitonov and K. B. Efetov, Phys. Rev. B \textbf{77}, 045116 (2008).
\bibitem{PRB-1989-Mobius-w}A. M\"{o}bius, Phys. Rev. B \textbf{40}, 4194 (1989).
\bibitem{CRSS-2017-Mobius}A. M\"{o}bius, Crit. Rev. Solid State, DOI: 10.1080/10408436.2017.1370575.
\bibitem{Wiley-1962-MacDonald}D. K. C. MacDonald, Thermoelectricity: an Introduction to the Principles (Wiley, New York, London, 1962).

\bibitem{JAP-2004-Li-ITO}Z. Q. Li and J. J. Lin, J. Appl. Phys. \textbf{95}, 5918 (2004).
\bibitem{JAP-2010-Wu-ITO}C. Y. Wu, T. V. Thanh, Y. F. Chen, J. K. Lee, and J. J. Lin, J. Appl. Phys. \textbf{108}, 123708 (2010).
\bibitem{APL-2014-Lang-FTO}W. J. Lang, and Z. Q. Li, Appl. Phys. Lett. \textbf{105}, 042110 (2014).
\bibitem{Nanotec-2009-Chiu}S. P. Chiu,  H. F. Chung, Y. H. Lin, J. J. Kai, F. R. Chen and J. J.Lin, Nanotechnology \textbf{20}, 105203 (2009).
\bibitem{ASS-2012-Liu}X. D. Liu, J. Liu, S. Chen, and Z. Q. Li, Appl. Surf. Sci. \textbf{263}, 486 (2012).

\bibitem{PRB-2013-Preissler}N. Preissler, O. Bierwagen, A. T. Ramu, and J. S. Speck, Phys. Rev. B \textbf{88}, 085305 (2013).
\bibitem{PRB-2008-Fuchs}F. Fuchs and F. Bechstedt, Phys. Rev. B \textbf{77}, 155107 (2008).
\bibitem{PSSB-1966-Tauc}J. Tauc, R. Grigorovici, and A. Vancu, Phys. Status Solidi \textbf{15}, 627 (1966).
\bibitem{Prentice-1971-Pankove}J. I. Pankove, Optical processes in semiconductors (Prentice-Hall, Englewood Cliffs, 1971).
\bibitem{Wiley-1972-Barnad}R. D. Barnard, Thermoelectricity in Metals and Alloys (Wiley, New York, 1972).
\end{thebibliography}
\end{document}